\documentclass[12pt]{JHEP3}

\usepackage{epsfig}
\usepackage{latexsym}
\usepackage{amssymb, amsmath, amsopn, amsthm}
\def\journal#1&#2(#3){\unskip, \sl #1\ \bf #2 \rm(19#3) }
\def\andjournal#1&#2(#3){\sl #1~\bf #2 \rm (19#3) }

\def\a{\alpha}
\def\g{\gamma}

\def\frac#1#2{{#1\over#2}}

\def\inbar{\,\vrule height1.5ex width.4pt depth0pt}
\def\IC{\relax\hbox{$\inbar\kern-.3em{\rm C}$}}
\def\IR{\relax{\rm I\kern-.18em R}}
\def\IP{\relax{\rm I\kern-.18em P}}

\catcode`\@=11
\def\slash#1{\mathord{\mathpalette\c@ncel{#1}}}
\def\lam{\lambda}

\def\underrel#1\over#2{\mathrel{\mathop{\kern\z@#1}\limits_{#2}}}

\catcode`\@=12

\def\det{{\rm det}}

\def \cosh{{\rm cosh}}

\def\det{{\rm det}}
\def\exp{{\rm exp}}

\newcommand{\p}{\partial}
\newcommand{\be}{\begin{equation}}
\newcommand{\ee}{\end{equation}}

\newcommand{\go}{g_{\rm o}}

\def\[{[}
\def\]{]}

\def\comment#1{ }

\def\draftnote#1{\ifdraft{\baselineskip2ex
                 \vbox{\kern1em\hrule\hbox{\vrule\kern1em\vbox{\kern1ex
                 \noindent \underbar{NOTE}: #1
             \vskip1ex}\kern1em\vrule}\hrule}}\fi}
\def\internote#1{\ifinter{\baselineskip2ex
                 \vbox{\kern1em\hrule\hbox{\vrule\kern1em\vbox{\kern1ex
                 \noindent \underbar{Internal Note}: #1
             \vskip1ex}\kern1em\vrule}\hrule}}\fi}
\def\inbar{\hskip.3em\vrule height1.5ex width.4pt depth0pt}
\def\IC{\relax{\inbar\kern-.3em{\rm C}}}
\def\IN{\relax{\rm I\kern-.16em N}}
\def\IQ{\relax\hbox{$\inbar$\kern-.3em{\rm Q}}}
\def\IZ{\relax{\rm Z\kern-.8em Z}}
\def\be{\begin{equation}}
\def\ee{\end{equation}}
\def\bea{\begin{eqnarray}}
\def\eea{\end{eqnarray}}

\title{A solution to the 4-tachyon off-shell amplitude in cubic string field theory}

\author{Valentina Forini\\Dipartimento di Fisica and I.N.F.N. Gruppo
Collegato di Trento,
Universit\`a di Trento, 38050 Povo (Trento). Italia.
\email{E-mail:forini@science.unitn.it}
\thanks{Work supported by INFN of Italy.}}

\author{Gianluca Grignani\\Dipartimento di Fisica and Sezione I.N.F.N.,
Universit\`a di Perugia, Via A. Pascoli I-06123, Perugia, Italia.
\email{E-mail:grignani$@$pg.infn.it}
\thanks{Work supported by INFN and MIUR of Italy and partially supported by INFN-MIT ``Bruno Rossi"
Exchange Program.}}

\author{Giuseppe Nardelli\\Dipartimento di Fisica and I.N.F.N. Gruppo
Collegato di Trento,
Universit\`a di Trento, 38050 Povo (Trento). Italia.
\email{E-mail:nardelli@science.unitn.it}
\thanks{Work supported by INFN of Italy.}}

\abstract{We derive an analytic series solution of the elliptic
equations providing the 4-tachyon  off-shell amplitude in cubic
string field theory (CSFT). From such a solution we compute the
exact coefficient of the quartic effective action relevant for time
dependent solutions and we derive the exact coefficient of the
quartic tachyon coupling. The rolling tachyon solution expressed as
a series of exponentials $e^t$ is studied both using
level-truncation computations and the exact 4-tachyon amplitude. The
results for the level truncated coefficients are shown to converge
to those derived using the exact string amplitude. The agreement
with previous work on the subject, both on the quartic tachyon
coupling and on the CSFT rolling tachyon, is an excellent test for
the accuracy of our off-shell solution. }

\preprint{}


\begin{document}

\section{Introduction and Conclusions}
\label{intro}

Cubic string field theory (CSFT)~\cite{Witten:1985cc} has played an
important role in recent years in describing the dynamics of the
open bosonic string tachyon. Both the unstable vacuum and the "true"
vacuum where the tachyon has condensed have been shown to be
well-defined states in CSFT~\cite{{Sen:2004nf}}. Tachyon
condensation is an off-shell process and string field theory is the
natural setting for its analysis.  The condensation process should
be described by the solutions of the equation of motion of the
tachyon effective action which can be constructed perturbatively in
string field theory. The tachyon effective action in fact can be
derived from the off-shell tachyon amplitudes, which can be computed
in various ways in string field theory. Following the classification
of ref.\cite{Taylor:2004rh}, there are four possible approaches for
computing off-shell amplitudes that we briefly describe here since
three of them have been used in this paper.
\begin{description}
\item[a)] Field theory
approach

The string field contains an infinite number of component fields,
whose number grows exponentially with the mass level $L$. In this
approach one can approximate the calculations by truncating the
string field up to some fixed level $L$~\cite{Kostelecky:1988ta},
for this reason it is called ``level truncation on fields". For
example one can construct the CSFT lagrangian by means of a
truncated string field up to some level $L$ and then compute the
cubic terms for each of the field components at the desired level.
 From this classical action one can then derive the tree level
effective action for some field component ($e.g.$ the tachyon) by
integrating out all the other ones through the solution of their
equations of motion. We shall use this procedure in Sections
\ref{sec:potential}-\ref{sec:rolling} to derive the perturbative
tachyon effective action.

\item[b)] Conformal mapping

With this method Giddings~\cite{Giddings:1986iy} reproduced the
on-shell Veneziano amplitude directly from Witten's CSFT. He gave an
explicit conformal map that takes the Riemann surfaces defined by
the Witten diagrams to the standard disc with four tachyon vertex
operators on the boundary. Following Giddings' procedure and with
some additional analysis -related to the oscillator method in c)-
Samuel~\cite{Samuel:1987uu} and Sloan~\cite{Sloan:1987bf} computed
the off-shell Veneziano amplitude. This procedure allows in
principle the calculation of any amplitude~\cite{Samuel:1989fe}.
Amplitudes computed using this method are exact, although numerical
approximations are necessary to get concrete numbers for them. We
shall solve Samuel's equations to derive, from the 4-tachyon off
shell amplitude, some very accurate numerical approximations of the
quartic coupling of the tachyon potential and of a time dependent
solution of CSFT.

\item[c)] Oscillator method

Perturbative amplitudes can be directly evaluated using the
oscillator representation of the vertices and propagators in CSFT.
The vertex and the propagator can be written completely in terms of
squeezed states~\cite{Kostelecky:2000hz}, \emph{i.e.} in terms of
exponentials of quadratic forms in the oscillators creating and
annihilating operators. In this way the complete set of amplitudes
associated with a Feynmann diagram results in an integral over the
internal  momenta that can be evaluated using standard squeezed
techniques (see Section \ref{sec:off-shell}). Any perturbative
amplitude is then given in a closed-form expression containing
infinite-dimensional Neumann matrices. While no analytical way is
known at present to exactly calculate such expressions, one can
evaluate the amplitudes to a high degree of precision truncating the
Neumann matrices to finite size~\cite{Taylor:2002bq}. This means
truncate the levels of oscillators in the string states which are
considered, this is the reason for which this method is known as
``level truncation on oscillators". Rather then having to include a
number of fields which grows exponentially in the level, with this
procedure one simply needs to evaluate quantities, as the
determinant of the Neumann matrices, whose size grows linearly in
the truncation level. A specific example of this method is given in
Appendix \ref{app:B}.

\item[d)] Moyal string field theory

In this alternative formulation of SFT the string joining star
product is identified with the Moyal product. Calculations performed
using this method reproduce directly the expressions for the
off-shell amplitudes as for example the $3$-point and $4$-point
tachyon amplitudes~\cite{Bars:2003sm}. Some numerical
results~\cite{Bars:2002yj} achieved with this procedure are
comparable to those obtained using the methods (a)-(c).

\end{description}

In this paper we mainly focus on the four tachyon amplitude which we
evaluate both by solving explicitly Samuel's elliptic equations for
the off-shell factor (method (b)) and by level truncation (methods
(a) and (c)). In particular we have obtained a new series solution
for the off-shell factor introduced by Samuel~\cite{Samuel:1987uu},
which, at variance with the one found in~\cite{Bars:2003sm},
provides the off-shell factor in terms of the original coordinates
used in~\cite{Samuel:1987uu}. From this solution we shall then
extract off-shell information both on the non-perturbative stable
vacuum and on the tachyon dynamics.

As a test for the solution we shall first improve the numerical
approximation for the evaluation of the exact quartic self-coupling
$c_4$ in the tachyon potential. This was computed for the first time
in~\cite{Kostelecky:1988ta} and repeated to a higher degree of
precision in~\cite{Taylor:2002bq}. Our results provides $c_4$ with a
precision that goes up to the ninth significative digit and is in
complete agreement with the extrapolations of
ref.~\cite{Beccaria:2003ak}.

As a second application we shall improve the CSFT time-dependent
solution given in~\cite{Coletti:2005zj} as a sum in powers of
$e^{t}$.

Since Sen's seminal paper on the rolling tachyon~\cite{Sen:2002nu},
much work has been devoted to the study of time dependent solutions
in string
theory~\cite{Sen:2002vv,Larsen:2002wc,Moeller:2002vx,Fujita:2003ex,Forini:2005bs,Coletti:2005zj}.
The setting is realized by considering a system of unstable
$D$-branes which decays in time as the tachyon field rolls down from
the maximum of the potential towards the stable minimum. A review on
previous work on this problem is given in~\cite{{Sen:2004nf}}.

The dynamics of a rolling tachyon has been studied in various
frameworks. The physical picture emerging from the boundary states,
RG flow and boundary string field theory (BSFT)
approaches~\cite{Minahan:2002if,Larsen:2002wc,Gaberdiel:2004na} is
quite clear. The tachyon rolls from the perturbative to the true
vacuum, which is reached in an infinite time. The same physical
picture can also be obtained following other approaches, among them
the analysis involving DBI-type
actions~\cite{Sen:2003tm,Kutasov:2003er,Garousi:2003ai,Fotopoulos:2003yt},
S-branes and time-like Liouville
theory~\cite{Strominger:2002pc,Gutperle:2003xf,Leblond:2003db,
Strominger:2003fn,Schomerus:2003vv}, matrix
models~\cite{McGreevy:2003kb,Klebanov:2003km,Constable:2003rc,
McGreevy:2003ep,Takayanagi:2003sm,Douglas:2003up}, cubic superstring
 field theory~\cite{Aref'eva:2003qu}, vacuum
SFT~\cite{Fujita:2004ha,Bonora:2004kf,Bonora:2005kz} and fermionic
boundary CFT~\cite{Lee:2005ge}.

CSFT instead fails to provide a meaningful description of the
rolling tachyon dynamics. At the lowest order, the $(0,0)$, in the
level truncation scheme one considers only the tachyon field and the
cubic string field theory action becomes
\begin{equation}
S=\frac{1}{\go^2}\int d^{26}x\left( \frac12\, \phi(x)\,(\Box+1)\,
\phi(x) -\frac13 \lambda
\left(\lambda^{(1/3)\Box}\phi(x)\right)^3\right) , \label{taction}
\end{equation}
where the coupling $\lam$ has the value $
\lambda=3^{9/2}/2^6=2.19213 $.  Considering spatially homogeneous
profiles of the form $\phi(t)$, where $t$ is time, the equation of
motion derived from (\ref{taction}) is
\begin{equation}
(\p_t^2-1) \phi(t)+\lambda^{1-\p_t^2/3} \left(\lambda^{-\p_t^2/3}
\phi(t)\right)^2=0. \label{eom}
\end{equation}
This equation was studied
in~\cite{Moeller:2002vx,Fujita:2003ex,Forini:2005bs,Coletti:2005zj}.
In~\cite{Forini:2005bs} it was found an almost exact well behaved
 solution of this equation for $\lambda<1$.
The solution has interesting analytical properties and is remarkably
simple. The ``evolution" of the solution to different values of
$\lambda$ is driven
 by a diffusion equation
which makes Eq.(\ref{eom}) local with respect to the time variable
$t$.
 The analytic
continuation of this solution to the physical value
$\lambda=3^{9/2}/2^6$ can be performed for any time $t$ with the
exception of a single point, $t=0$, where the solution is not
analytic.
 The profile can be
expressed in terms of a series in powers of $e^t$ for $t<0$ and in
powers of $e^{-t}$ for $t>0$ and in this way it is well-behaved
except at the origin where it has a cusp. Alternatively,  one can extend
to positive $t$ the solution in powers of $e^t$ and the solution
presents  ever-growing oscillations.
 In any case the tachyon always
rolls well past the minimum of the potential then turns around.
Solutions with ever-growing oscillations have been found also in
refs.~\cite{Moeller:2002vx,Fujita:2003ex,Coletti:2005zj}.
In~\cite{Coletti:2005zj}, in particular, a systematic
level-truncation analysis was carried out for a trajectory $\phi(t)$
expressed as a power series in $e^t$. It was also shown that the
non-local field redefinition, which takes the CSFT action to the
BSFT action~\cite{Coletti:2004ri}, also maps the wildly oscillating
CSFT solution to the well-behaved BSFT exponential solution.
Increasing the level of truncation in CSFT or the number of terms
retained in the tachyon effective action leads to a well defined
trajectory at least up to some upper bound in $t$, $t=t_b$. In fact,
if the position of the first turnaround points, that the solution
exhibits for $t>0$, tends to stabilize as the truncation level $L$
of the effective action increases, the expansion in powers of $e^t$
for $t>0$ would be justified at least up to those
points~\cite{Coletti:2005zj}. For the first turnaround points, the
leading terms in the CSFT solution are those with small powers of
$e^t$. Consequently, the very accurate value of the 4-tachyon
amplitude that we have found in this paper improves the solution of
ref.~\cite{Coletti:2005zj}, at least up to the first extrema of the
trajectory.

The trajectories $\phi(t)$, obtained by computing the $\phi^4$ term
in the effective action exactly and  the terms up to $\phi^7$ in an
$L=2$ approximation, show that indeed the position of the first
turnaround point does not change significantly with the improvement
in  the $\phi^4$ term. This suggests the possibility that this value
actually has the physical meaning of inversion point. The second
turnaround point instead changes position and amplitude compared to
the one found in~\cite{Coletti:2005zj}. The inclusion of higher
order terms in the lagrangian however does not produce significative
changes, so that the trajectory seems again to stabilize. Thus we
confirm that for $t>0$, the tachyon does not roll towards the stable
non-perturbative minimum of the potential and that the  qualitative
behavior of wild oscillations is reproduced even if the amplitudes
at the turnaround points beyond the first are  sensibly diminished.

The solutions of the 4-tachyon off-shell amplitude that we have
found therefore is a very useful tool for providing precise tests of
CSFT. The agreement with previous work on the subject, both on the
quartic tachyon coupling and on the CSFT rolling tachyon, is an
excellent test for the accuracy of our off-shell solution.

As for the DBI tachyon action, it would be instructive to study the
cubic tachyonic action on a curved background and, in particular, in
a Friedmann–Robertson–Walker (FRW) spacetime. It would be
interesting to see if the coupling of the free theory to a
Friedman-Robertson-Walker metric~\cite{Gibbons:2002md}, and the
consequent inclusion of a Hubble friction term, might lead from the
classical solution with ever-growing oscillations to damped
oscillations around the stable minimum of the potential well. Cubic
string field theory might then open interesting perspectives in
tachyon cosmology \cite{Calcagni:2005xc}.

The paper is organized as follows. In Section~\ref{sec:off-shell} we
review the derivation of the off-shell four tachyon amplitude
following ref.~\cite{Samuel:1987uu}. Explicit formulas for the
Neumann coefficients involved in the oscillator formalism are
reported in Appendix~\ref{app:A}. A brief review of the level
truncation method is also given and a specific example is provided
in Appendix \ref{app:B}. In Section \ref{sec:kappa} we develop the
tools needed to perform the computations of Sections
\ref{sec:potential}-\ref{sec:rolling}. A solution to the elliptic
equations defining the off-shell amplitude is derived, obtaining a
useful expansion of $\kappa(x)$ in powers of the Koba-Nielsen
variable $x$. This analysis improves the accuracy in the evaluation
of the quartic coupling of the tachyon potential, which is performed
in Section \ref{sec:potential}. Finally, in Section
\ref{sec:rolling} we use the exact four-point amplitude
 to study the first few coefficients of the
rolling tachyon solution expressed as a sum of exponentials $e^{n
t}$, and we compare the corresponding solution
 with the ones obtained in the
level truncation scheme.

Our calculations were performed using the symbolic manipulation
program \emph{Mathematica}.

\section{Off-shell 4-tachyon amplitude}
\label{sec:off-shell}

The first step in computing the off-shell four tachyon amplitude in
CSFT is to construct the Feynman diagrams directly from the cubic
interaction vertex. Four-point amplitudes involve one propagator and
two vertices. After gauge fixing, we use the Feynman-Siegel gauge,
the propagator becomes $b_0/L_0$ where $L_0$ is the Virasoro
generator for the intermediate state including ghosts
\begin{equation}
L_0=p\cdot p-1+\sum_{n=1}^{\infty}\left(\alpha_{-n}\cdot\alpha_n+n
b_{-n}c_{n}+n c_{-n}b_n\right)\,\,.
\end{equation}
Writing the propagator
$$
\frac{b_0}{L_0}=b_0 \int_0^\infty d T e^{-T L_0}\;,
$$
$e^{-T L_0}$ inserts a world-sheet strip of length $T$ into the
amplitude.

\subsection{Conformal mapping: on-shell amplitude }
\label{sec:conformal}

\begin{figure}
\begin{center}
\includegraphics[scale=0.8]{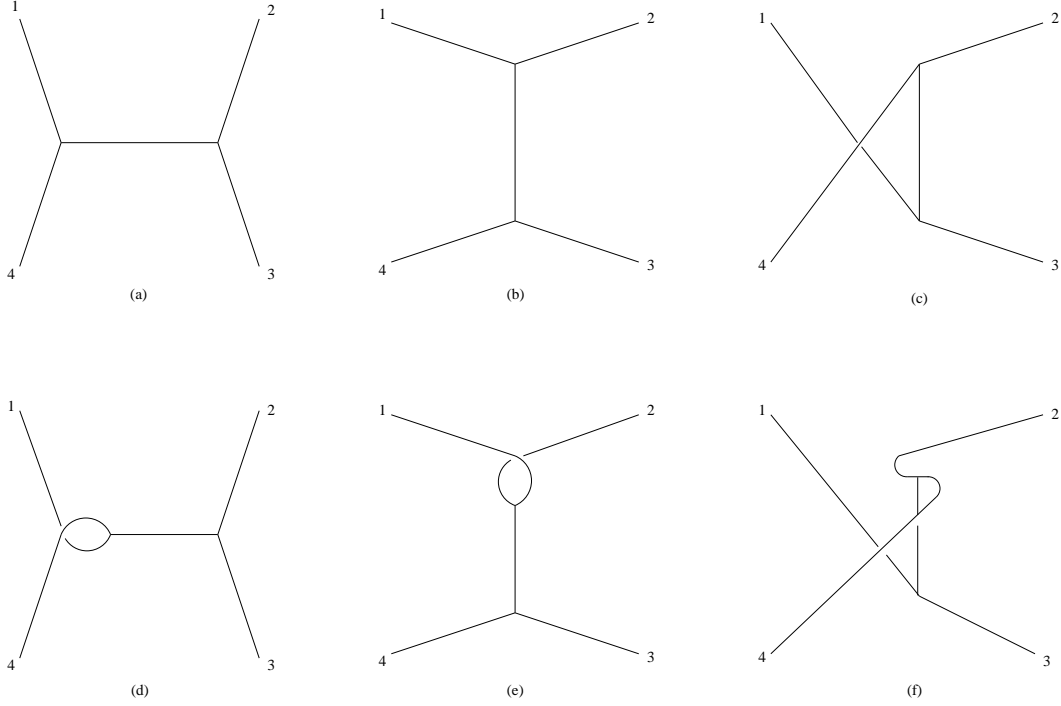}
\caption{The relevant Feynman diagrams for the four particles
scattering.} \label{feynman}
\end{center}
\end{figure}

A closed analytical expression for the off-shell four tachyon
amplitude in CSFT~\cite{Witten:1985cc} was derived
in~\cite{Samuel:1987uu} by following Gidding's analysis of the
on-shell Veneziano amplitude~\cite{Giddings:1986iy}. Giddings gave
an explicit conformal map that takes the Riemann surfaces defined by
the Witten diagrams to the standard disc with four tachyon vertex
operators on the boundary. This conformal map is defined in terms of
four parameters $\alpha,~\beta,~\gamma,~\delta$.
 The four parameters are not independent variables. They satisfy the relations
\begin{equation}
\a\beta=1\quad\quad\gamma\delta=1
\label{parameters}
\end{equation}
and
\begin{equation}
\frac{1}{2}=\Lambda_0(\theta_1,k)-\Lambda_0(\theta_2,k)\;,
\label{constraint}
\end{equation}
where $\Lambda_0(\theta,k)$ is defined by
\begin{equation}
\Lambda_0(\theta,k)=\frac{2}{\pi}\left(E(k)F(\theta,k^\prime)
+K(k)E(\theta,k^\prime)-K(k)F(\theta,k^\prime)\right)\;\;.
\label{lambda}
\end{equation}
In (\ref{lambda}) $K(k)$ and $E(k)$ are complete
elliptic functions of the first and second kinds, $F(\theta,k)$ is
the incomplete elliptic integral of the first kind (we follow the
notation of ref.\cite{Gradshteyn}). The parameters $\theta_1$,
$\theta_2$, $k$ and $k^\prime$  satisfy
\begin{eqnarray}
k^2 = \frac{\gamma^2}{\delta^2} & \hspace{1in} &  k'^2 = 1-k^2\\
\sin^2 \theta_1 =  \frac{\beta^2}{\beta^2 + \gamma^2} &&
 \sin^2\theta_2 = \frac{\alpha^2}{ \alpha^2 + \gamma^2}  \,.
\end{eqnarray}
By convention $\beta >\alpha$ and $\delta >\gamma$. Because of
(\ref{parameters}) and (\ref{constraint}) only one variable is
independent. By convention this is taken to be $\a$, that is related
to $T$, the lenght of the intermediate strip, by
\begin{equation}
\label{alphaT} \frac{T}{2}=K(k^\prime)\left[Z(\theta_2,k^\prime)-
Z(\theta_1,k^\prime)\right]
\end{equation}
where $Z(\theta,k)$ is defined through the ordinary elliptic
functions
\begin{equation}
Z(\theta,k)=K(k)E(\theta,k)-E(k)F(\theta,k)\;.
\end{equation}
The parameter $\a$ is finally related to the Koba-Nielsen variable
$x$ through
\begin{equation}
x=\left(\frac{(1-\alpha^2)}{(1+\alpha^2)}\right)^2\
,~~~~~~~~~~~~~~\alpha=\sqrt{\frac{1-\sqrt{x}}{1+\sqrt{x}}}\ .
\label{alphaofx}
\end{equation}
Using this conformal map Giddings managed to derive the Veneziano
amplitude from CSFT. Because of the cubic vertex, in CSFT there are
six relevant Feynman diagrams for four particles processes
(fig.\ref{feynman}). The contribution from the graph (a) in
fig.\ref{feynman} was computed in~\cite{Giddings:1986iy} to be
\begin{equation}
A_s(p_1,p_2,p_3,p_4)=\int_{\a_0}^0 d\a\, 2
A_G~\frac{dT}{d\alpha}~(\beta-\alpha)^{2(p_1\cdot p_2+p_3\cdot
p_4)}(\beta+\alpha)^{2(p_1\cdot p_3+p_2\cdot p_4)}
(2\alpha)^{2(p_2\cdot p_3)}(2\beta)^{2(p_1\cdot p_4)}
\label{amplimap}
\end{equation}
where the integration limits $\a_0=\sqrt{2}-1$ and $\a=0$ correspond
to $T=0$ and $T=\infty$ respectively , $2A_G$ is the ghost
contribution and is given by
\begin{equation}
2 A_G=
8\frac{1}{2\pi}\sqrt{\alpha^2+\gamma^2}\sqrt{\beta^2+\gamma^2}(\beta^2-\alpha^2)K(\gamma^2)
\end{equation}
and the Jacobian factor almost cancels the ghost factor
\begin{equation}
\frac{dT}{d\alpha}=-\frac{4(\beta^2-\alpha^2)}{\alpha A_G}~~~~~.
\end{equation}

\subsection{Oscillator method: off-shell amplitude }
\label{sec:conformal}

Samuel derived a perturbative off-shell string
amplitude~\cite{Samuel:1987uu} directly from string field theory by
requiring that it reproduces Gidding's result (\ref{amplimap}) when
the momenta are set on-shell. We now briefly review Samuel results.

Let
\begin{equation}
\frac{g}{2}\langle{V^{(3)}_{41I}}|\langle{V^{(3)}_{23J}}|b_0 e^{-T
L_0}|V^{(2)}_{IJ}\rangle=\,\langle{V^{(4)}_{1234}}|
\end{equation}
denote the vertex function associated with the graph (a) in
fig.\ref{feynman}, where the subscripts $1,2,3,4, I$ and $J$
indicate Fock-space labels. The full contribution to the diagram is
\begin{equation}
\int_0^\infty dT
\langle{V^{(4)}_{1234}}||\Psi^{(4)}_4\,\rangle|\Psi^{(3)}_3
\rangle|\Psi^{(2)}_2\rangle|\Psi^{(1)}_1\rangle \label{1a}
\end{equation}
where $|\Psi^{(r)}_r\rangle$ is the Fock-space representation of the
external states.
 The explicit oscillator representations of
$\langle{V^{(2)}}|$ and $\langle{V^{(3)}}|$
\begin{equation}
\langle{V^{(2)}_{12}}|=\int d^{26}  p \;
\langle{p}|^{(1)}\otimes\langle{p}|^{(2)} \left( c_0^{(1)}+
c_0^{(2)} \right) e^{-\sum_{n=1}^\infty (-1)^n \left[a_n^{(1)}\cdot
a_n^{(2)} + c_n^{(1)}  b_n^{(2)}+c_n^{(2)}  b_n^{(1)}\right]}
\end{equation}
\begin{eqnarray}
\langle{V^{(3)}_{123}}|=&&\int d^{26}  p_1 d^{26}  p_2 d^{26}  p_3\;
\delta(p_1+p_2+p_3)\langle{p_1}|^{(1)}c_0^{(1)}
\otimes\langle{p_2}|^{(2)}c_0^{(2)}\otimes\langle{p_3}|^{(3)}c_0^{(3)}\cdot\cr
&&\cdot e^{-\frac{1}{2}\sum_{r,s=1}^3 \left[a_m^{(r)}V_{mn}^{rs}
a_n^{(s)} +2 a_m^{(r)}V^{r s}_{m 0} p^{(s)} +p^{(r)}V^{r s}_{0 0}
p^{(s)}-2c_m^{(r)}X_{mn}^{rs} b_n^{(s)}\right]}
\end{eqnarray}
show that all the terms in (\ref{1a}) are given in terms of
exponentials of quadratic expressions in the oscillators. Using
standard squeezed state techniques~\cite{Kostelecky:2000hz},
closed-form expressions can be given for any perturbative amplitude.
In the case of the four tachyon amplitude corresponding to the first
diagram of fig.\ref{feynman}, this procedure gives\footnote{We
follow the notation of refs.\cite{Taylor:2002bq,Coletti:2003ai}.}

\begin{equation}
A_4(p_1,p_2,p_3,p_4)\!=\!\frac{\lambda_c^2
g^2}{2}\,\delta(\textstyle{\sum_i} p_i)\!\int_0^\infty dT\,e^T\,\det
\left(\textstyle{\frac{1-(\tilde{X}^{11})^2}
{1-(\tilde{V}^{11})^2}}\right)e^{-\frac{1}{2}p_i Q^{i
j}p_j}\label{amplineumann}
\end{equation}
where $\lambda_c$ is a constant related to the
Neumann coefficient for the three tachyon vertex, $\lambda_c=e^{3
V_{00}^{11}}=\frac{3^{9/2}}{2^6}$ . In this formula $\tilde{V}^{11}$
and $\tilde{X}^{11}$ are defined by
\begin{equation}
\tilde{V}^{11}_{m n}=e^{-\frac{(m+n) }{2}T}V^{11}_{m
n}\;\;\;\;\;\;\;\;\;\tilde{X}^{11}_{m
n}=e^{-\frac{(m+n)}{2}T}X^{11}_{m n}
\end{equation}
where $V^{rs}$ and $X^{rs}$ are infinite-dimensional matrices
\begin{align}
V^{rs}&=\left(
\begin{array}{ccccc}
V^{rs}_{11} & V^{rs}_{12}  &\dots & V^{rs}_{m n}&\dots   \\
V^{rs}_{21} & V^{rs}_{22}  &\dots & V^{rs}_{m+1, n}&\dots    \\
\dots & \dots & \dots & \dots & \dots
\end{array}\right), &
X^{rs}&=\left(
\begin{array}{ccccc}
X^{rs}_{11} & X^{rs}_{12}  &\dots & X^{rs}_{m n}&\dots   \\
X^{rs}_{21} & X^{rs}_{22}  &\dots & X^{rs}_{m+1, n}&\dots    \\
\dots & \dots & \dots & \dots & \dots
\end{array}\right)
\label{V&X}
\end{align}
whose elements are matter and ghost Neumann coefficients of the
cubic string field theory vertex, for which exact expressions are
given in the Appendix \ref{app:A}. $Q^{i j}$ are defined as
\begin{eqnarray}
Q^{i j}=&&V^{i I}_{0m}\left(\textstyle{\frac{1}
{1-(\tilde{V}^{11})^2}}\right)_{m n}\tilde{V}^{11}_{n p}V^{I j}_{p
0}+V_{00}^{11}-T(2-\delta_{ij}) \;\;\;\;\;\;i,j=1,2
\;\hbox{or}\;i,j=3,4\cr Q^{i j}=&&-V^{i
I}_{0m}\left(\textstyle{\frac{1} {1-(\tilde{V}^{11})^2}}\right)_{m
n}\,C \,\tilde{V}^{11}_{n p}V^{I j}_{p 0} \quad\;\;\;\;\;\;i=1,2
\;\hbox{and}\;j=3,4\;\hbox{or}\; i=3,4 \;\hbox{and}\;j=1,2\cr &&
 \label{Qij}
\end{eqnarray}
where $m,n,p\geq1$, $C=\delta_{mn}(-1)^n$ and the sum over $I$
denotes a sum over the intermediate states.

The two expressions (\ref{amplimap}) and (\ref{amplineumann}) should
both represent the contribution to the four tachyon amplitude coming
from the diagram (a) in fig.\ref{feynman} when the momenta are
on-shell. To relate them in the proper way, a general procedure was
developed in~\cite{Samuel:1987uu} for computing the functions
$Q^{ij}$ appearing in (\ref{amplineumann}) from the Giddings map,
giving
\begin{align}
\nonumber Q^{11} &= Q^{44}=\ln\a-\ln \kappa
 , & Q^{22}=Q^{33} &=-\ln\a-\ln \kappa\\
\nonumber Q^{12} &= Q^{21}=-\ln|\a-\beta|
 , & Q^{13}=Q^{31} &=-\ln(\a+\beta )\\
\nonumber Q^{14} &= Q^{41}=-\ln(2\beta)
 , & Q^{23}=Q^{32} &=-\ln(2\a)\\
 Q^{24}&=Q^{42}=-\ln(\a+\beta)
 , & Q^{34}=Q^{43} &=-\ln|\a-\beta|
\label{Qijmapped}
\end{align}
where $\kappa$ is given as an integral
\begin{equation}
\ln(\kappa)=-2\a \textstyle{\frac{(\beta^2-\alpha^2)}{\sqrt{
(\alpha^2+\g^2)(\a^2+\delta^2)}}}\int_1^\infty \!\!dw \ln(w-1)
\frac{d}{d
w}\left(\textstyle{\frac{\sqrt{(w^2+\alpha^2\gamma^2)(w^2+\a^2
\delta^2)}}{(w+1)(\beta^2 w^2-\alpha^2)}}\right)\;\;.
\label{logkappa}
\end{equation}
As already noticed, although $\alpha,\beta,\gamma,\delta$ all appear
in the above equation, there is only one independent variable, so
that the function $\kappa$ in (\ref{mapneumann}) is actually a
function of $\a$. The substitution of (\ref{Qijmapped}) in
(\ref{amplineumann}) leads to the following formula
\begin{eqnarray}
A_4(p_1,p_2,p_3,p_4)&=&\lambda_c^2\frac{g^2}{2}\int_{\a_0}^0 d\a
\frac{dT}{d\a}e^T\det \left(\textstyle{\frac{1-(\tilde{X}^{11})^2}
{1-(\tilde{V}^{11})^2}}\right)\left[\kappa(\a)\right]^{\sum_{i=1}^4
p^2_i} \left(\a\right)^{-(p_1^2+p_4^2)+p_2^2+p_3^2}\cr
&&|\a-\beta|^{2(p_1\cdot p_2+p_3\cdotp_4)}(\beta+\alpha)^{2(p_1\cdot
p_3+p_2\cdotp_4)} (2\alpha)^{2(p_2\cdot p_3)}(2\beta)^{2(p_1\cdot
p_4)} \label{mapneumann}
\end{eqnarray}
Comparing the two expressions (\ref{amplimap}) and
(\ref{mapneumann}) on shell ($p^2_i=1$), one can see that the
momentum dependence matches and for the momentum independent part
the following identity holds
\begin{equation}
\lambda_c^2 \left(\frac{dT}{d\a}\right) e^T \det
\left(\textstyle{\frac{1-(\tilde{X}^{11})^2}
{1-(\tilde{V}^{11})^2}}\right)=2 A_g \frac{dT}{d\a}
\frac{1}{\left[\kappa(\a)\right]^4}\;\;. \label{match}
\end{equation}

By trading the variable $\a$ for the Koba-Nielsen variable $x$
through (\ref{alphaofx}) in (\ref{mapneumann}), the contribution
from the first graph in fig.\ref{feynman} becomes
\begin{equation}
A_4(p_1,p_2,p_3,p_4)=\frac{g^2}{2}\int_{\frac{1}{2}}^1 dx\; x^{p_1
\cdot p_2+p_3\cdot
p_4}(1-x)^{(p_1+p_4)^2-2}\left(\frac{\kappa(x)}{2}\right)^{\sum_{i=1}^4
p^2_i-4} \label{amplitudemom}
\end{equation}

The remaining diagrams (b),(c),(d),(e),(f) of fig.\ref{feynman} can
be obtained from the first one by a suitable permutation of the
string labels, \emph{i.e.} by permuting the momenta in
(\ref{amplitudemom}), and the total four-point tachyon amplitude is
the sum of these six contributions. Notice that the Veneziano
amplitude is exactly reproduced when $p^2_i=1$ in
(\ref{amplitudemom}) and the additional factor containing
$\kappa(x)$ goes to $1$.

\subsection{Level truncation}
\label{sec:level truncation}

 The infinite-dimensional matrices
(\ref{V&X}) appearing in the final expression for a given diagram
are expressed in terms of the Neumann coefficients of Witten's
vertex. The level truncation method we use in this paper consists in
the truncation on the level of oscillators associated with the
Neumann coefficients. This procedure is somewhat different from the
original method of level truncation~\cite{Kostelecky:1988ta} (method
a) section~\ref{intro}), in which one calculates the SFT action by
only including in the string field expansion contributions up to a
fixed total oscillator level. While the latter approach involves
computations with a number of fields that grows exponentially in the
level, in the former one has to calculate the determinant of some
matrices whose size grows linearly in the truncation level.

Let us explicitly remind the procedure~\cite{Coletti:2003ai} in the
case of a tree diagram with four external fields as
(\ref{amplineumann}), in which there is a single internal propagator
with Schwinger parameter $T$. One starts with a suitable change of
coordinates in (\ref{amplineumann})
\begin{equation}
\sigma=e^{-T}
\label{coord}
\end{equation}
 then expands in powers of $\sigma$, so getting an
expression of the form
\begin{equation}
\int_0^1\frac{d\sigma}{\sigma^2}\sigma^{p^2}\sum_{n=0}^{\infty}c_n(p_i)\sigma^n
=\sum_{n=0}^{\infty}\frac{c_n(p^i)}{p^2+n-1}\;\;, \label{expansion}
\end{equation}
where $p=p_1+p_2=p_3+p_4$ represents the momentum of the
intermediate state. The poles $p^2=1-n$ in (\ref{expansion}) clearly
correspond to the contributions of intermediate particles as the
tachyon ($n=0$), the gauge field ($n=1$) and all the other open
string massive fields. Truncate all the matrices to size $L\times L$
means to truncate the sum in (\ref{expansion}) to $n=L$, thus
imposing a limit on the mass of the intermediate states.

The analysis can be simplified by noting that in the four point
amplitude the contributions of odd level fields cancel between $s$
and $t$ channels so that only even levels in the truncation,
\emph{i.e.} only even powers of $\sigma$ in the expansion
(\ref{expansion}), need to be considered. An explicit example of the
procedure above explained is given in Appendix \ref{app:B}, where
the four tachyon amplitude at level $L=2$ is derived in the
time-dependent case.

\section{Solution for the function $\kappa(x)$}
\label{sec:kappa}

As shown in the previous section the off-shell 4-point string
amplitudes are completely determined once the function $\kappa(x)$
defined by (\ref{logkappa}) is known. To determine the function
$\kappa(\a,\g)$ we have first to solve eq.(\ref{constraint}) for one
of the two variables in terms of the other, so that the function
$\kappa$ will be a function of only one of the two $\alpha$ or
$\gamma$. Since the four point amplitude is written in terms of an
integral over $x$, which is easily related to $\alpha$ through
(\ref{alphaofx}), it would be more natural to solve for $\gamma$ as
a function of $\alpha$ then the opposite. The solution can be found
numerically and for $\gamma$ as a function of $x$ is given by the
solid line in fig.\ref{g+g}. $\gamma$ goes from 0 to 1 while $x$
goes from 1 to 1/2 and $\alpha$ goes from 0 to $\sqrt{2}-1$. To
check for the accuracy of the solution, we have found two different
expansions: 1) A power series in $\alpha$ which gives $\gamma$ in a
neighbor of $0$ and can be inverted so as to give $\alpha$ as a
function of $\gamma$ around 0. 2) An expansion of $\alpha$ around
$\sqrt{2}-1$ as an expansion in $1-\gamma$, this series cannot be
inverted due to the presence of terms of the type $(1-\g)^m
\log(1-\gamma)^n$. We have found a general procedure to obtain as
many terms as necessary in both expansions and the function
$\alpha(\gamma)$ can be determined in the whole range $0\le\gamma\le
1$. As we shall show in fact the two series for $\alpha(\gamma)$
overlap in an extended interval that goes from $\gamma\sim 0.6$ to
$\gamma\sim 0.7$.

\subsection{$\gamma$ and $\alpha$ around $0$}
\label{sec:around0}

By using the integral representations of the elliptic
functions~\cite{Gradshteyn} it is possible to write the equation
(\ref{constraint}) in a useful form
\begin{equation}
E(\gamma^2)\int_{\alpha\gamma}^{\gamma/\alpha}dt\frac{1}{\sqrt{t^2+
\gamma^4}\sqrt{1+t^2}}-
(1-\gamma^4)K(\gamma^2)\int_{\alpha\gamma}^{\gamma/\alpha}dt
\frac{1}{\sqrt{t^2+\gamma^4}(\sqrt{1+t^2})^3}=\frac{\pi}{4}\;
\label{elliptic}
\end{equation}
To expand (\ref{elliptic}) for small $\gamma$ and $\alpha$  we have
to divide the integration region into three intervals in such a way
that the square roots in the denominators of (\ref{elliptic}) can be
consistently expanded and the integrals in $t$ performed. For
example consider the integral in the first term of (\ref{elliptic}),
it can be rewritten as \bea
&&\int_{\alpha\gamma}^{\gamma/\alpha}dt\frac{1}{\sqrt{t^2+\gamma^4}\sqrt{1+t^2}}=\cr
&&\int_{\alpha\gamma}^{\gamma^2}
dt\frac{1}{\gamma^2\sqrt{1+\frac{t^2}{\gamma^4}}\sqrt{1+t^2}}+\int^1_{\gamma^2}
dt\frac{1}{t
\sqrt{1+\frac{\gamma^4}{t^2}}\sqrt{1+t^2}}+\int_1^{\frac{\gamma}{\alpha}}
dt\frac{1}{t^2 \sqrt{1+\frac{\gamma^4}{t^2}}\sqrt{1+\frac{1}{t^2}}}
\cr&& \eea

In each integral of the rhs the integration domain is contained in
the convergence radius of the Taylor expansions of the square roots
containing $\g$, so that they can be safely expanded and the
integrals in $t$ performed.

\begin{figure}
\begin{center}
\includegraphics[scale=0.9]{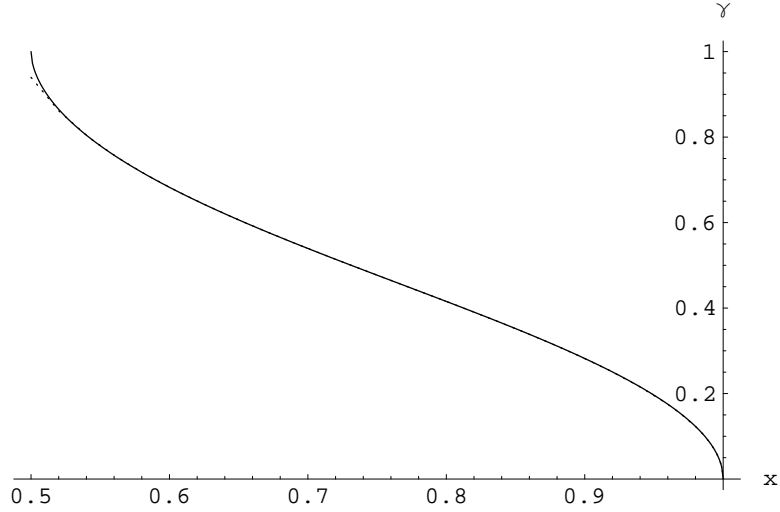}
\caption{Plots of $\gamma(x)$: the solid line is the numerical
solution of the elliptic equation, the dashed line is the power
series.} \label{g+g}
\end{center}
\end{figure}

With this procedure one gets the following equation equivalent to
(\ref{elliptic}) \bea
&&E(\gamma^2)\sum_{n,k=0}^{\infty}{\frac{\Gamma(\frac{1}{2})^2}
{\Gamma(\frac{1}{2}-n)\Gamma(\frac{1}{2}-k)n!
k!}}\cr&&\left\{\frac{2}{2n+2k+1}
\left[\gamma^{4k}-\left(\frac{\a}{\g}\right)^{2n+1}\left(\a\g\right)^{2k}\right]
+\left(1-\delta_{k n}\right)\frac{\g^{4n}-\g^{4k}}{2k-2n}-\delta_{k
n}\g^{4n}\ln{\g^2}\right\}\cr&&
-(1-\g^4)K(\g^2)\sum_{n,k=0}^{\infty}{\frac{\Gamma(\frac{1}{2})\Gamma(-\frac{1}{2})}
{\Gamma(\frac{1}{2}-n)\Gamma(-\frac{1}{2}-k)n!
k!}}\cr&&\left\{\frac{1}{2n+2k+1}
\left[\gamma^{4k}-\left(\frac{\a}{\g}\right)^{2n+1}\left(\a\g\right)^{2k}\right]
+\left(1-\delta_{k n}\right)\frac{\g^{4n}-\g^{4k}}{2k-2n}-\delta_{k
n}\g^{4n}\ln{\g^2}\right.\cr && \left.+\frac{1}{2n+2k+3}
\left[\gamma^{4n}-\left(\frac{\a}{\g}\right)^{2k+3}\left(\a\g\right)^{2n}\right]\right\}=\frac{\pi}{4}
\label{ellipticexp} \eea The series containing $\ln\gamma^2$ can be
resummed, the first gives $\frac{2}{\pi} K(\gamma^2)$ the second
$\frac{2}{\pi(1-\gamma^4)}E(\gamma^2)$. Hence these terms cancel and
 $\ln\gamma^2$ actually disappears from the equation. As a
consequence one can write $\gamma$  as a power series in $\a$ whose
coefficients are determined requiring that eq.(\ref{ellipticexp}) is
satisfied. $\g$ turns out to contain only the powers $\a^{4 n+1}$,
 $n\in\mathbb{N}$. We have determined the first 12 terms of this
series to get a very good approximation for $\g$ in an extended
neighbor of zero (in which sense it is an extended neighbor will be
clarified later) \bea
&&\g=\sqrt{3}\a\left(1+5\a^4+\frac{1041}{16}\a^8+\frac{38719}{32}\a^{12}
+\frac{109062913}{4096}\a^{16}
+\frac{5278728465}{8192}\a^{20}+\right.\cr&&\left.
\frac{2172202186251}{131072}\a^{24}+\frac{116561474500179}{262144}\a^{28}+
\frac{3303689940814193505}{268435456}\a^{32}+\right.\cr
&&\left.\frac{187301165958864015157}{536870912}\a^{36}+
\frac{86571446884950765378149}{8589934592}\a^{40}+\right.\cr&&\left.
\frac{5078927050639748451791733}{17179869184}\a^{44}+\,O(\a^{48})\right)
\label{gammaofalpha} \eea Any higher  order in (\ref{gammaofalpha})
can be in principle computed from (\ref{ellipticexp}). Using
(\ref{alphaofx}) we can plot $\g$ as a function of $x$ and compare
it to the graph obtained from the numerical solution of
eq.(\ref{elliptic}). As it is clear from fig.\ref{g+g} $\g(x)$ has
in $x=1/2$ a vertical tangent, thus showing the presence of a branch
point which cannot be gotten from a power series of the form
(\ref{gammaofalpha}). Nevertheless (\ref{gammaofalpha}) gives a very
good approximation for $\g(x)$ except in a small neighbor of
$x=1/2$. In particular the agreement between the values of $\g$
obtained from the series (\ref{gammaofalpha}) and the numerical
values is on the 15-th significative digit for $0.8\le x\le 1$,
where the series (\ref{gammaofalpha}) is expected to give exact
results, thus providing a precision test for the accuracy of the
numerical solution.
Moreover, the expansion (\ref{gammaofalpha}) can be iteratively
inverted to give a series for $\a$ as a function of $\g$ \bea
&&\a=\frac{\g}{\sqrt{3}}\left(1-\frac{5}{9}\g^4+\frac{959}{1296}\g^8-
\frac{10993}{7776}\g^{12}+\frac{83359631}{26873856}\g^{16}-
\frac{3579242677}{483729408}\g^{20}+\right.\cr&&\left.\frac{1297273056905}{69657034752}\g^{24}
-\frac{6783253984031}{139314069504}\g^{28}
+\frac{168109910408625655}{1283918464548864}\g^{32}-\right.\cr&&\left.
\frac{24949101849547687507}{69331597085638656}\g^{36}
+\frac{10046339553062261150885}{9983749980331966464}\g^{40}
-\right.\cr&&\left.\frac{512861712698825472832315}{179707499645975396352}\g^{44}+O(\g^{48})\right)
\label{alphaofgamma} \eea By plugging the expansion
(\ref{gammaofalpha}) in (\ref{logkappa}) and using (\ref{alphaofx}),
the corresponding expansion for $\kappa(\a)$ can be found by means
of numerical integration \bea \kappa(\a)=\frac{8}{3\sqrt{3}}\;\exp
&&\!\!\!\!\left[-2.5 \,\a^4 - 7.1562 \,\a^8 - 75.927
\,\a^{12}-1238.7 \,\a^{16}-24301\,\a^{20}\right.\cr &&
\left.-531290\,\a^{24}-1.2489\cdot 10^7 \,\a^{28} -3.0923\cdot10^8\,
\a^{32}\right.\cr && \left.- 7.9627 \cdot10^9 \,\a^{36}-2.1140
\cdot10^{11} \,\a^{40}-5.7517 \cdot10^{12} \,\a^{44}\right]
+O(\a^{48})\ .\cr &&\label{kappasint} \eea

\subsection{$\gamma$ around 1 and $\alpha$ around $\sqrt{2}-1$}
\label{sec:around1}

Around $x=1/2$, \emph{i.e} $\a=\sqrt{2}-1$ and $\g=1$, it is
possible to obtain only $x$ (or $\a$) as a function of $\g$ and not
the opposite. Such an expansion can be obtained by first expanding
eq.(\ref{elliptic}) around $\g=1$ and then looking for an expansion
of $\a$ in terms of powers of $1-\g$ and $\ln(1-\g)$ \bea
&&\a=\sqrt{2}-1+a_1(1-\g)+a_2(1-\g)^2+ \dots +b_1(1-\g)\ln(1-\g)+\cr
&&b_2(1-\g)^2\ln(1-\g)+\dots+
c_1(1-\g)(\ln(1-\g))^2+c_2(1-\g)^2(\ln(1-\g))^2+\dots \cr
&&\label{ansatz} \eea The coefficients in (\ref{ansatz}) are
determined by requiring that (\ref{elliptic}) is satisfied. We
provide here directly the expansion of $x$ as a function of $1-\g$
up to the ninth order
\begin{eqnarray}
&&x=\frac{1}{2}+\frac{1}{8}(1-\g)^2\left[1-2\log\left(\textstyle{\frac{1-\g}{4}}\right)\right]-
\frac{1}{4}(1-\g)^3\log\left(\textstyle{\frac{1-\g}{4}}\right)-\cr&&
\frac{1}{16}(1-\g)^4\left[1+3\log\left(\textstyle{\frac{1-\g}{4}}\right)\right]-
\frac{1}{96}(1-\g)^5\left[7+12\log\left(\textstyle{\frac{1-\g}{4}}\right)\right]+\cr
&&\frac{1}{1536}(1-\g)^6\left[-97-108\log\left(\textstyle{\frac{1-\g}{4}}\right)-
24\log^2\left(\textstyle{\frac{1-\g}{4}}\right)
+64\log^3\left(\textstyle{\frac{1-\g}{4}}\right)\right]-\cr&&
\frac{1}{2560}(1-\g)^7\left[119+100\log\left(\textstyle{\frac{1-\g}{4}}\right)
-40\log^2\left(\textstyle{\frac{1-\g}{4}}\right)-
320\log^3\left(\textstyle{\frac{1-\g}{4}}\right)\right]+\cr&&
\frac{1}{10240}(1-\g)^8\left[-321-60\log\left(\textstyle{\frac{1-\g}{4}}\right)+
1240\log^2\left(\textstyle{\frac{1-\g}{4}}\right)+
2240\log^3\left(\textstyle{\frac{1-\g}{4}}\right)\right]+\cr&&
\frac{1}{107520}(1-\g)^9\left[-1871+5740 \log
   \left(\textstyle{\frac{1-\g}{4}}\right)+29120 \log
   ^2\left(\textstyle{\frac{1-\g}{4}}\right)
   +31360 \log
   ^3\left(\textstyle{\frac{1-\g}{4}}\right)\right]+\dots\cr
   &&
\end{eqnarray}

\begin{figure}
\begin{center}
\includegraphics[scale=0.8]{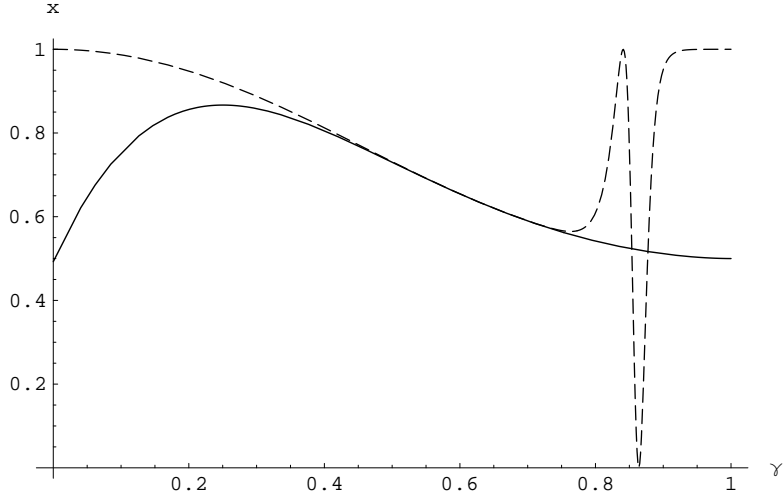}
\caption{Plots of $x(\gamma)$: the dashed line gives the expansion
of $x(\g)$ which holds in a neighbor of $\g=1$, the solid line gives
the expansion of $x(\g)$ around $\g=0$. } \label{xofg}
\end{center}
\end{figure}

From (\ref{alphaofgamma}) one can easily get $x$ as a function of
$\g$ in the region $x\sim 1$ ($\g\sim 0$) so that $x(\g)$ can be
obtained for the whole range $1/2\le x\le 1$. The two expansions in
fact overlap in a long range for $0.3\le\g\le 0.7$ as it is shown in
fig.\ref{xofg}. They have an excellent agreement up to the 13-th
significative digit  for $0.6\le\g\le 0.7$.

\section{Coefficient of the Quartic Tachyon Potential}
\label{sec:potential}

The static tachyon potential has the form\footnote{We follow the
notation of refs.\cite{Taylor:2000ek,Taylor:2002bq}.}
\begin{equation}
V_T=\frac{1}{2}\phi^2-g\, k\,\, \phi^3+ g^2 k^2\, c_4 \phi^4+\dots
\label{potential}
\end{equation}
where $g$ is the string coupling constant and $k =
\frac{3^{7/2}}{2^7}$.

The four point tachyon potential  is obtained from the off-shell
four tachyon amplitude by setting to zero the external momenta and
by explicitly subtracting out the term with the tachyon on the
internal line. The amplitude is the sum of the six Feynman diagrams
shown in fig.\ref{feynman}, the first of which gives the
contribution (\ref{amplitudemom}) that can be usefully rewritten in
terms of the Mandelstam variables
\begin{equation}
A4(s,t,u)=\frac{g^2 \lambda_c^2}{2} \int_{\frac{1}{2}}^1 \;dx\;
x^{\frac{t-s-u}{2}}(1-x)^{s-2}\left(\frac{\kappa(x)}{2}\right)^{t+s+u-4}\;\;.
\label{amplitude} \end{equation} To get explicitly the first diagram
contribution to the amplitude one can set $t=u=0$ in
(\ref{amplitude}), $A_4$ can then be defined through an analitical
continuation of (\ref{amplitude}) to the region $s\leq 1$. This can
be achieved by adding and subtracting the pole in $x=1$ in the
integrand of (\ref{amplitude}) \bea &&\int_{\frac{1}{2}}^1
dx\;x^{-\frac{s}{2}}(1-x)^{s-2}
\left(\frac{\kappa(x)}{2}\right)^{s-4}=\int_{\frac{1}{2}}^1
dx\;x^{-\frac{s}{2}}(1-x)^{s-2}
\left[\left(\frac{\kappa(x)}{2}\right)^{s-4}-\left(\frac{\kappa(1)}{2}\right)^{s-4}\right]\cr
&&+\left(\frac{\kappa(1)}{2}\right)^{s-4}\int_{\frac{1}{2}}^1\;dx\;x^{-\frac{s}{2}}(1-x)^{s-2}\;\;.
\label{analcont} \eea where the first integral is now well defined
in $s=0$. When $\textit{Re}[s]>1$ the last integral in
(\ref{analcont}) gives
$$
\frac{2^{s-2}}{\sqrt{\pi}}\Gamma(1-\frac{s}{2})\Gamma(\frac{s}{2}-\frac{1}{2})
+\frac{2^{2-\frac{s}{2}}}{s-2}
~_2F_1\left(1,2-s;2-\frac{s}{2};-1\right)
$$
 that
has a well defined limit for $s\to 0$, so that the four point
tachyon potential can be written
 \be A4(0,0,0)=\frac{g^2
\lambda_c^2}{2} \left[\int_{\frac{1}{2}}^1 dx
\left(\left(\frac{2}{\kappa(x)}\right)^4-\left(\frac{2}{\kappa(1)}\right)^4\right)
(1-x)^{-2}-\frac{3}{2}\left(\frac{2}{\kappa(1)}\right)^4\right]
\label{pot} \ee As already pointed out, the function $\kappa(x)$ in
(\ref{pot}) can be evaluated numerically in the whole interval
$\frac{1}{2}<x<1$, by using the numerical solution of
eq.(\ref{elliptic}) graphed in fig.\ref{gammax}. The integrand in
(\ref{pot}) is regular at $x=1$, as can be easily checked by
studying the behavior of (\ref{kappasint}) in a neighbor of $\a=0$.
However, problems are expected in the numerical evaluation of the
integral in a neighbor of $x=1$ due to the product of a pole times a
zero. To circumvent possible computational problems we divided the
interval $\frac{1}{2}<x<1$ into two parts . For $x\in
[\frac{1}{2},0.95]$ we used numerical evaluation of $\kappa(x)$, by
plugging the numerical solutions of (\ref{elliptic}) in
(\ref{logkappa}). For $x\in[0.95,1]$ we used the analitical
expression obtained substituting (\ref{alphaofx}) in
(\ref{kappasint}). By summing the two contributions we have found
the value $A4(0,0,0)=-\frac{g^2}{2}2.94497480(2)$. To get the the
quartic term of the tachyon effective potential we have to
subtract~\cite{Kostelecky:1988ta} from (\ref{pot}) the contribution
from the internal tachyon line \be
A4_t(s,t,u)=\frac{g^2}{2}\lam_c^{2-s-\frac{t+u}{3}}\frac{1}{s-1}
\label{at} \ee evaluated at $s=t=u=0$. Each graph in
fig.\ref{feynman} contributes equally,  so that for the quartic
tachyon coupling one eventually gets
\begin{equation}
g^2 k^2 c_4=\frac{6 }{4!}
\left[A4(0,0,0)-A4_t(0,0,0)\right]=\frac{6}{4!}\frac{g^2  }{
2}(-2.94497480(2) + \frac{ 3^9}{2^{12}})=  \frac{g^2 }{4!}
\;5.5813353(1) \label{c4}
\end{equation}
where the factor $1/4!$ is required to recover the units
of~\cite{Taylor:2000ek,Taylor:2002bq}. The numerical evaluation of
the coefficient $c_4$ from the exact four tachyon amplitude was
given in~\cite{Samuel:1987uu} to an accuracy of $1\%$, $c_4\approx
1.75(2)$, and in~\cite{Taylor:2002bq} to an accuracy of $0.1\%$,
$c_4\approx 1.742(1)$. We have repeated this calculation to an
higher degree of precision, and the result (\ref{c4}) gives
\begin{equation}
c_4\approx1.74220008(3)\;. \label{c4final}
\end{equation}
This coefficient was calculated using the level truncation scheme up
to level $L=20$ in~\cite{Taylor:2000ek}, and improved up to level
$L=28$ in~\cite{Beccaria:2003ak}, thus obtaining
$c_{4,L=28}\simeq-1.70028$, with a discrepancy of $2.4\%$ with
respect to (\ref{c4final}). In the same paper, a procedure to
extrapolate the known level truncated results and predict the
asymptotic $L\to\infty$ value for $c_4$ was described, giving an
extimated value $c_{4,L\to\infty}=1.7422006(9)$ that agrees within
the $10^{-7}$ of accuracy with our exact result (\ref{c4final}).

\section{The rolling tachyon in cubic string field theory}
\label{sec:rolling}

As a second application of the formalism developed in Section
\ref{sec:off-shell}, we discuss some properties of the rolling
tachyon solutions in CSFT. This problem has been faced analytically
in~\cite{Forini:2005bs} at the $(0,0)$ level, and  numerically
in~\cite{Moeller:2002vx,Fujita:2003ex,Coletti:2005zj}. In
particular, a level truncated analysis of the tachyon dynamics was
carried out in~\cite{Coletti:2005zj} for a perturbative solution
given as a sum of exponentials of the form
\begin{equation}\label{rollsum}
\phi(t)=\sum_{n>0} a_n e^{n t}\;\;.
\end{equation}
 The solution and all its derivatives satisfy the boundary condition
$\phi\to 0$ as $t\to -\infty$. The coefficients in (\ref{rollsum})
can be determined by perturbatively solving the CSFT equation of
motion. For such a profile the $\phi^{n+1}$ term in the tachyon
effective action contributes only to the coefficients $a_k$ with
$k\ge n$. Since in the CSFT tachyon effective action
\begin{equation}
S[\phi]=\sum_n \frac{g^{n-2}}{n!}\int\prod_{i=1}^n(2\pi
dk_i)\delta(\sum_i k_i) \phi(k_1) \dots
\phi(k_n)A_n(k_1,\ldots,k_n)\;\;\; \label{CSFT}
\end{equation}
the coefficients $A_2$ and $A_3$ are exactly known,
\begin{equation}
A_2(k_1,k_2)=1- k_1 k_2 \ ,\quad\quad \quad \quad
A_3(k_1,k_2,k_3)=-2\left(\frac{3
\sqrt{3}}{4}\right)^{3+k_1^2+k_2^2+k_3^2}\ , \label{A2A3}
\end{equation}
the first two coefficients in (\ref{rollsum}) are exact and can be
normalized as $a_1=1$, $a_2=-64/(243 \sqrt{3})$.

In~\cite{Coletti:2005zj} an $L=2$ approximation  was
explicitly provided for the coefficients $a_3\ldots a_6$ in the sum (\ref{rollsum})
\begin{equation}
\phi(t)\cong e^t-\frac{64}{243\sqrt{3}}e^{2t}+0.002187
\,e^{3t}-3.9258\, 10^{-6}\,e^{4t}+4.9407\, 10^{-10}\,e^{5t}-6.3227\,
10^{-12}e^{6t} \label{rollingtaylor}
\end{equation}

For  negative $t$ Eq.(\ref{rollingtaylor}) describes the rolling of
the tachyon off the unstable maximum along the potential. The
physical interpretation for positive $t$ is more problematic. The
truncated expansion (\ref{rollingtaylor}) is a solution only up to
some upper bound $t=t_b$ which increases by increasing the number of
terms one includes in the sum. Consequently, the asymptotic behavior
of the solution for large positive  $t$ cannot be extrapolated from
Eq.(\ref{rollingtaylor}), being the sum alternate the asymptotic
behavior would simply be $\pm \infty$ depending on the order $n$ at
which one truncates the sum (\ref{rollsum}).

Before exponentially exploding $\phi (t)$ presents an oscillatory
behavior with increasing amplitudes that makes the rolling tachyon
dynamics in the framework of CSFT for positive $t$ difficult to
interpret. In ref.~\cite{Coletti:2005zj}, however, it was shown that
the trajectory $\phi(t)$ is well-defined. Increasing both the level
of truncation and the number of terms retained in the power series
(\ref{rollsum}) leads to a convergent value of $\phi(t)$ for any
fixed $t$ with $t<t_b$. If the position of the first turnaround
points, that the solution exhibits for $t>0$, tends to stabilize as
the truncation level $L $ of the effective actions increases, the
expansion (\ref{rollingtaylor}) for $t>0$ would be justified at
least up to those points. The trajectories $\phi(t)$, obtained by
computing the $\phi^4$ term in the effective  action up to $L=16$,
show that indeed the position of the first two turnaround points
seems to stabilize~\cite{Coletti:2005zj}. For $t>0$, the tachyon
does not roll towards the stable non-perturbative minimum of the
potential.

We shall now study how this solution is modified by using the exact
value of the 4-tachyon term in the effective action for homogeneous
time dependent profiles. The exact value of the coefficient $a_3$
can be obtained by computing integrals of the type
(\ref{amplitudemom}), that in the time-dependent case read
\begin{equation}
A_4(p_1,p_2,p_3,p_4)=\frac{g^2}{2}\int_{\frac{1}{2}}^1 dx\; x^{-p_1
\cdot p_2-p_3\cdot
p_4}(1-x)^{-(p_1+p_4)^2-2}\left(\frac{\kappa(x)}{2}\right)^{-\sum_{i=1}^4
p^2_i-4} \label{amplitudetime}
\end{equation}
To get the equations of motion the function $A_4$ in (\ref{CSFT})
has to be evaluated for imaginary integer values of the field modes
so that (\ref{amplitudetime}) is regular and does not need any
analytical continuation. In the evaluation of $a_3$, the relevant
integral (\ref{amplitudetime})
 over the Kobe-Nielsen variable is
\begin{equation}\label{A4a3}
A_4(-i,-i,-i,3i) =\frac{g^2}{2}\int_{\frac{1}{2}}^1 dx\;
x^{-2}(1-x)^{2}\left(\frac{\kappa(x)}{2}\right)^{8}
\end{equation}
Summing all the diagrams in  fig.\ref{feynman} and subtracting the
corresponding contributions coming from the internal tachyon line,
$A_{4t}=2^{29}/3^{22}$, we get $a_3=0.00241475435(3)$. This value,
which is exact, can  be compared with the corresponding ones
obtained through the level truncation approximation.
\begin{table}[htp]
\begin{center}
\begin{tabular} {|l|l | l | l | l | }
\hline \hline
Level & $a_3$ &   $a_4$ &  $a_5$&  $a_6$  \\
\hline \hline
2  &  0.002187797562  & $ -3.7830611 \,10^{-6}$  & $ 4.1448524\,10^{-9}$ &  $ -4.7728992\,10^{-13}$ \\
\hline
4  &  0.002245884478 & $ -4.3957017\,10^{-6}$ & $ 4.6338501\,10^{-9}$& $ -5.4000742\,10^{-13}$ \\
\hline
6  & 0.002281097505  &  $ -4.5437634\,10^{-6}$  &  $ 4.7480437\,10^{-9}$ &  $ -6.2618454\,10^{-13}$ \\
\hline
8  &  0.002304369408  & $ -4.6509193\,10^{-6}$  &  $ 4.8933743\,10^{-9}$  &  $-6.7366480\,10^{-13}$\\
\hline
10  &  0.002320816678  &  $ -4.7282645\,10^{-6}$  &  $ 4.9938778\,10^{-9}$ &  $ -6.9213556\,10^{-13}$ \\
\hline
12  & 0.002333033369 &  $-4.7867688\,10^{-6}$& $5.0729134\,10^{-9}$ & $-7.0850857\,10^{-13}$ \\
\hline
14  & 0.002340032469  &  $-4.8250629\,10^{-6}$  &$5.1236425\,10^{-9}$ & $-7.2267875\,10^{-13}$ \\
\hline
16  & 0.002342489534  & $-4.8443632\,10^{-6}$  & $5.1338898\,10^{-9}$ &  $-7.3568697\,10^{-13}$ \\
\hline \hline $\hbox{Exact} A_4$ &  0.00241475435(3)  &
$-5.205903(1) \,10^{-6}$  &  $ 5.692641(2)\,10^{-9}$
& $-8.338132(4)\,10^{-13}$  \\
\hline \hline
\end{tabular}
\end{center}
\caption{\footnotesize First few coefficients $a_n$ of the
time-dependent solution $\sum_n a_n e^{n t}$ at various levels of
truncation, when only the contribution from the quartic term in the
effective action is considered in the EOM. In the last row the exact
four tachyon amplitude is used for the calculations.}
\label{coefficients}
\end{table}
The first column of Table 1 shows the sequence of the first
approximate values of the $a_3(L)$ coefficients up to $L=14$. The
level sequence is perfectly consistent with the exact value given in
the last row (first column), which should then be considered as the
limit $a_3(L\to \infty)$.

The amplitude (\ref{amplitudemom}) can be used to improve the
accuracy of the remaining coefficients $a_n$, $n\ge 4$. The exact
evaluation  of $a_4$ would require the knowledge of $A_5 (p_1,\ldots
, p_5)$, for which an expression analog to (\ref{amplitudemom}) is
not known. However, when solving the CSFT equation of motion, one
can easily see that the dominant contribution to $a_4$ comes from
the lower order amplitudes $A_2 (p_1, p_2)$, $A_3 (p_1, p_2, p_3)$,
$A_4 (p_1, p_2, p_3, p_4)$.  Therefore, for a precise evaluation of
$a_4$ seems more relevant to know these lower order amplitudes
exactly, rather than $A_5 (p_1,\ldots , p_5)$ approximate in levels.
The remaining columns in Table 1 give the behavior of the
coefficients $a_4$, $a_5$, $a_6$ for increasing levels of
truncation, when only the contribution from the quartic term in the
effective action is considered in the equations of motion. The last
row gives the corresponding value obtained from the exact amplitude
(\ref{amplitudemom}) ({\it i.e.}
 limit $L\to \infty$).
As can be seen from Table 1, for any fixed $L$, $ |a_n (L)|< |a_n
(L\to \infty)| $. Notice that the same property holds also  in the
calculation of the coefficient of the quartic tachyon potential.
Indeed, up to $L=28$~\cite{Beccaria:2003ak}, $|c_{4,L}|<|c_4|$.
Moreover, for any fixed $n$, the sequence $(a_n (L+2)-a_n (L))$ goes
like $C_n  a_n(L)/L$, $C_n$ being a constant, confirming the $1/L$
behavior of  the leading
correction~\cite{Taylor:2002fy,Beccaria:2003ak}. The results given
in the last row of Table 1 provide  the first few coefficients of
the trial solution (\ref{rollsum}).

We can now include in the computation of $a_4,\ a_5,\ a_6$ the $L=2$
truncated expressions for $A_5(p_1,\dots ,p_5),\ A_6(p_1,\dots
,p_6),\ A_7(p_1,\dots ,p_7)$. The numerical results are listed in
Table 2. The $L=2$ truncated $A_7(p_1,\dots ,p_7)$, however, gives a
contribution to $a_6$ which is not reliable, since increasing the
order of the effective action higher level field components become
more and more important. The inclusion of the $a_6$ coefficient, in
any case, does not  change the behavior of the solution around the
first two turnaround points. This is the region where we shall
mainly focus, only here the solution with the first few coefficients
is reliable.
\begin{table}[htp]
\begin{center}
\begin{tabular} {|l|l | l | l | }
\hline \hline
Effective action & $a_3$ &   $a_4$ &  $a_5$ \\
\hline \hline $A_4^{exact}$  &  0.00241475435(3)  & $-5.205903(1)
\,10^{-6}$
&  $ 5.692641(2)\,10^{-9}$  \\
\hline
$A_4^{exact}, A_5^{L=2}$  &  0.00241475435(3)  & $ -5.348643(1)10^{-6}$ & $ 3.231846(1)10^{-9}$ \\
\hline
$A_4^{exact}, A_5^{L=2} , A_6^{L=2}$  &  0.00241475435(3)  &  $ -5.348643(1)10^{-6}$  &  $ 2.0650063(5)10^{-9}$  \\
\hline \hline
\end{tabular}
\end{center}
\caption{\footnotesize First few coefficients $a_n$ of the
time-dependent solution $\sum_n a_n e^{n t}$. The first column
indicates which terms of the effective action are considered in the
EOM.} \label{coefficients2}
\end{table}

\begin{figure}
\begin{center}
\includegraphics[scale=0.8]{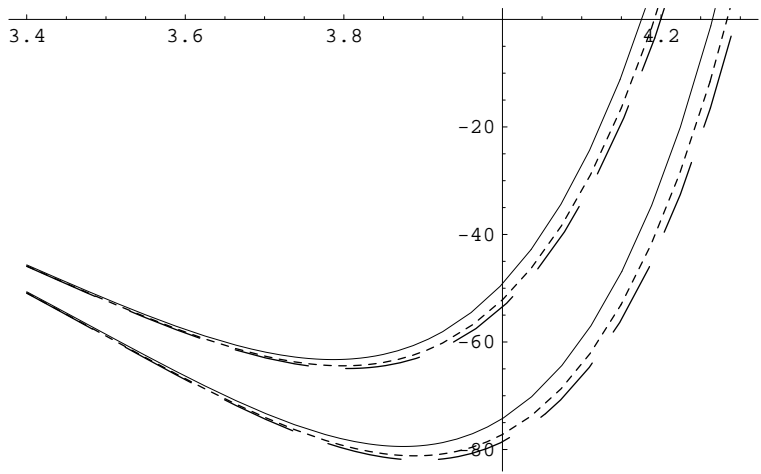}
\caption{Solution at the second turnaround point. The higher group
of trajectories is obtained by using the exact value for the
four-tachyon effective action (solid line) and adding to it the
level $L=2$ five (long dashed line) and six (dashed line) tachyon
effective action, the lower group by using only $L=2$ terms. The
solid line in the lower group represents the solution of ref.
\cite{Coletti:2005zj} up to the $e^{5t}$ power.} \label{solutions}
\end{center}
\end{figure}

\begin{figure}
\begin{center}
\includegraphics[scale=0.8]{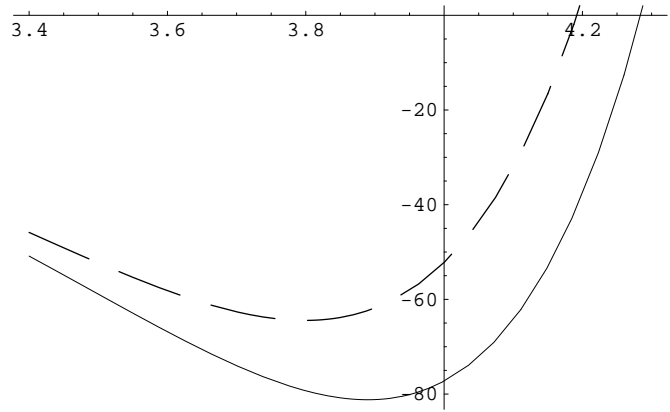}
\caption{Second turnaround point for the solution (solid line) given
in ref.~\cite{Coletti:2005zj} and the solution (dashed line)
 obtained using the exact $A_4$ and the level 2
$A_5$, $A_6$.} \label{turnaround}
\end{center}
\end{figure}

In fig.\ref{solutions} we show how the solution changes at the
second turnaround point by introducing higher order terms of the
effective action. The higher group of trajectories is obtained by
using the exact value for the four-tachyon effective action and
adding to it the level $L=2$ five and six  tachyon effective action,
the lower group by using only $L=2$ terms (the solid line in this
group represent the solution of ref. \cite{Coletti:2005zj} up to the
$e^{5t}$ power). As it is manifest from the figure the use of an
exact $A_4$ leads to a decreasing of the amplitude of the
oscillations by at least the $20\%$. This is however not enough to
change the qualitative behavior of the solutions which maintains
huge oscillations and does not provide a physically meaningful
picture. The best approximation we get is given by the solution
obtained using the exact $A_4$ and the level 2 $A_5$, $A_6$. It
reads
\begin{equation}
\phi(t)\cong e^t-\frac{64}{243\sqrt{3}}e^{2t}+0.00241475
\,e^{3t}-5.348643 10^{-6}\,e^{4t}+2.0650063 10^{-9}\,e^{5t}\
\label{rollingsam}
\end{equation}
and is plotted in fig.\ref{turnaround}  against the solution
(\ref{rollingtaylor}) of ref. \cite{Coletti:2005zj} up to the
coefficient of $e^{5t}$.

The solution (\ref{rollingsam}) can also be compared to the analytic
solution found in \cite{Forini:2005bs} at the  $(0,0)$ level that
reads, for $t<0$,
\begin{equation}
\phi(t)=-6 \lambda_c^{-\frac{5}{3}}\sum_{n=1}^\infty \left(\textstyle{-\frac{1}{6}}\right)^{n}
 n \lambda_c^{-\frac{4}{3}n^2 +3 n} e^{n t}\  \ ,\
\label{phinoi}
\end{equation}
 where $\lambda_c=3^{\frac{9}{2}}/{2^6}$. In \cite{Forini:2005bs}, a different expression was considered
for $t>0$. If however we  consider Eq.(\ref{phinoi}) also for
positive values of $t$, it can be conveniently compared to
(\ref{rollingsam}) and (\ref{rollingtaylor}). For $t<0$ all the
solutions overlap up to the $6$-th significative digit. For positive
$t$, all the solutions present the expected oscillatory behavior
with ever-growing amplitudes and have constant energy. In CSFT where
the action contains infinite derivatives the kinetic energy can be
negative and thus the tachyon can move to higher and higher heights
on the tachyon potential while conserving the total energy
\cite{Moeller:2002vx}. Whatever solution one chooses, the position
of the first extremum seems to be fixed at $t_1\sim 1.27$ with
amplitude $\phi(t_1)\sim 1.74$. In addition, such a position is
compatible within the $1 \%$ also with \cite{Moeller:2002vx}, where
an analog approximate solution was considered using the $\cosh nt$
basis. This suggests the idea that the first maximum could have a
physical meaning. Actually, since the solution describes the motion
of the tachyon rolling off its unstable maximum at $\phi=0$, the
naive energy conservation would confine the motion between $0\le
\phi(t)\le \phi_M$, where $\phi_M$ denotes the maximum value
attained by $\phi$ {\it i.e.}  is the naive inversion point defined
by the condition $V_{eff}[0]=V_{eff}[\phi_M]$ on the effective
tachyon potential $V_{eff}$. A natural interpretation for the first
maximum is therefore $\phi(t_1)\sim\phi_M$. Numerically, the value
$\phi_M\sim 1.7$ is in fact in a qualitative agreement with the
available data on the effective tachyon
potential~\cite{Moeller:2000xv}.

The other extrema, instead, do not have any clear physical meaning.
These  oscillations undergo wild ever-growing amplitudes, which,
however, depend quite significantly on the solution chosen. In
passing from (\ref{rollingtaylor}) to (\ref{rollingsam}), both
positions of the turnaround points and their amplitudes change. For
instance, as shown in fig.\ref{turnaround}, the amplitude of the
second turnaround point is lowered by a $20\%$ factor,  the third
one by an order of magnitude.

 In conclusion, it seems that up to the first turnaround point all
the solutions (\ref{rollingtaylor}), (\ref{rollingsam}),
(\ref{phinoi}), practically coincide. After the first turnaround
point, the wild oscillations with increasing amplitudes found in
refs.\cite{Moeller:2002vx,Coletti:2005zj} are confirmed. Although
the qualitative behavior is reproduced, the oscillations in
(\ref{rollingsam}) are  sensibly  reduced when compared to those in
ref.\cite{Coletti:2005zj}. Up to the second turnaround point, where
low powers of $e^t$ dominate, (\ref{rollingsam})  provides a more
accurate estimate for the trajectory of the rolling tachyon in CSFT.

\acknowledgments

We are grateful to E. Coletti, W. Taylor and I. Sigalov for useful
discussions.

\vspace*{0.2in}

\appendix

\section{Neumann Coefficients}
\label{app:A} \vspace*{0.1in}

Exact formulas for the Neumann coefficients $V^{rs}$ and $X^{rs}$
appearing in (\ref{V&X}) were computed
in~\cite{Gross:1986fk}\footnote{In some references signs and factors
in the Neumann coefficients may be slightly different. We follow
here the choices of~\cite{Taylor:2003gn}.}. The indices $r,s$ take
values from $1$-$3$ and indicate wich Fock space the oscillators act
in. The $3$-string coefficients $V^{rs}_{mn}$, $X^{rs}_{mn}$ are
given in terms of the $6$-string Neumann coefficients $N^{r, \pm
s}_{nm}$
\begin{eqnarray}
N^{r, \pm r}_{nm} & = & \left\{\begin{array}{l} \frac{1}{3 (n \pm
m)}  (-1)^n (A_nB_m \pm B_nA_m), \;\;\;\;\;
m + n\, {\rm even}, \;m \neq n\\
0, \;\;\;\;\; m + n\, {\rm odd}
\end{array} \right.\label{eq:n6}\\
N^{r, \pm (r + \sigma)}_{nm} & = & \left\{\begin{array}{l}
\frac{1}{6 (n \pm \sigma m)}  (-1)^{n + 1} (A_nB_m \pm \sigma
B_nA_m), \;\;\;\;\;
m + n\, {\rm even}, \;m \neq n\\
\sigma \frac{\sqrt{3}}{6 (n \pm \sigma m)} (A_nB_m \mp \sigma
B_nA_m), \;\;\;\;\; m + n\, {\rm odd}
\end{array}\right].
\label{6string}
\end{eqnarray}
where in $N^{r, \pm (r + \sigma)}$, $\sigma = \pm 1$, and $r
+\sigma$ is taken modulo 3 to be between 1 and 3. In (\ref{6string})
$A_n, B_n$ are defined for $n \geq 0$ through
\begin{eqnarray}
\left( \frac{1 + ix}{1-ix} \right)^{1/3}  & = & \sum_{n\, {\rm
even}} A_n x^n + i
\sum_{m\, {\rm odd}} A_m x^m  \label{eq:ab}\\
\left( \frac{1 + ix}{1-ix} \right)^{2/3}  & = & \sum_{n\, {\rm
even}} B_n x^n + i \sum_{m\, {\rm odd}} B_m x^m \,.  \nonumber
\end{eqnarray}
The 3-string matter Neumann coefficients $V^{rs}_{nm}$ are then
given by
\begin{eqnarray}
V^{rs}_{nm} & = &  -\sqrt{mn} (N^{r, s}_{nm} + N^{r, -s}_{nm}),
\;\;\;\;\; m \neq n,\, {\rm and}\, m, n \neq 0 \nonumber\\
V^{rr}_{nn} & = &  -\frac{1}{3}  \left[ 2 \sum_{k = 0}^{n}
(-1)^{n-k} A_k^2-(-1)^n-A_n^2 \right], \;\;\;\;\;
n \neq 0 \nonumber\\
V^{r, r + \sigma}_{nn} & = &\frac{1}{2} \left[ (-1)^n-V^{rr}_{nn}
             \right], \;\;\;\;\;  n \neq 0 \label{eq:n3}\\
V^{rs}_{0n}& = & -\sqrt{2n} \left( N^{r, s}_{0n} + N^{r, -s}_{0n}
\right), \;\;\;\;\; n \neq 0\nonumber\\
V^{rr}_{00} & = & \ln (27/16) \nonumber
\end{eqnarray}
The ghost Neumann coefficients $X^{rs}_{m n}, m\geq 0, n >0$ are
given by
\begin{eqnarray}
X^{rr}_{mn} & = & m \left( -N^{r, r}_{nm} + N^{ r, -r}_{nm} \right),
            \;\;\;\;\; n \neq
            m\nonumber\\
X^{r (r \pm 1)}_{mn} & = &  m \left(\pm N^{r, r \mp 1}_{nm} \mp N^{
r, - (r
            \mp 1)}_{nm} \right), \;\;\;\;\; n \neq
            m \label{eq:x3}\\
X^{rr}_{nn} & = &  \frac{1}{3} \left[ -(-1)^n-A_n^2 + 2 \sum_{k =
0}^{n}  (-1)^{n-k} A_k^2 -2
            (-1)^nA_nB_n \right] \nonumber\\
X^{r (r \pm 1)}_{nn} & = &
  -\frac{1}{2}(-1)^n -\frac{1}{2} X^{rr}_{nn}
\nonumber
\end{eqnarray}
The Neumann coefficients satisfy a cyclic symmetry under $r
\rightarrow r + 1, s \rightarrow s + 1$, corresponding to the
geometric symmetry of rotating the vertex. Furthermore, they are
symmetric under the exchange $r \leftrightarrow s, n \leftrightarrow
m$ and satisfy the twist symmetry associated with reflection of the
strings
\begin{eqnarray}
V^{rs}_{nm} & = &  (-1)^{n + m}V^{sr}_{nm}\\
X^{rs}_{nm} & = &  (-1)^{n + m}X^{sr}_{nm}\,. \nonumber
\end{eqnarray}

\section{Level truncation method}
\label{app:B}

As a specific example of the level truncation method explained in
Section \ref{sec:level truncation} let us derive explicitly the four
tachyon amplitude for $L=2$ in the time-dependent case. At this
level of truncation and with the change of coordinates
(\ref{coord}), the matrices $\tilde{V^{11}}$ and $\tilde{X^{11}}$ in
(\ref{amplineumann}) become
 the $2\times2$ matrices
\begin{align}
\tilde{V^{11}} &= \left(
  \begin{array}{cc}
    V^{11}_{11}\sigma & V^{11}_{12}\sigma^{\frac{3}{2}}  \\
    V^{11}_{21}\sigma^{\frac{3}{2}} & V^{11}_{22}\sigma^{2} \\
  \end{array}
\right), & \tilde{X^{11}}&= \left(
  \begin{array}{cc}
    X^{11}_{11}\sigma & X^{11}_{12}\sigma^{\frac{3}{2}}  \\
    X^{11}_{21}\sigma^{\frac{3}{2}} & X^{11}_{22}\sigma^{2} \\
  \end{array}
\right)
\end{align}
and analog forms for all the objects contained in (\ref{Qij}) may be
written. Expanding the determinant and the exponential in
(\ref{amplineumann}) in powers of $\sigma$ up to $\sigma^2$ one gets
\begin{eqnarray}
A_4(p_1,p_2,p_3,p_4)&=&\frac{\lambda_c^2
g^2}{2}\lambda_c^{\frac{2}{3}(\sum_{i=1}^4 p_i^2 +p_1\cdot
p_2+p_3\cdot p_4)}\,\delta(\sum_i p_i)\!\int_0^1
\frac{d\sigma}{\sigma^2}\sigma^{-\frac{1}{2}[(p_1+p_2)^2
+(p_3+p_4)^2]}\cr &&\left\{1-b_1(p_1-p_2)(p_3-p_4)\sigma+\frac{1}{2}
\left[b_2+ b_3\left((p_1 - p_2)^2 + (p_3 - p_4)^2\right)
\right.\right.\cr &&\left.\left.+b_4(p_1 - p_2)^2(p_3 - p_4)^2 +b_5
(p_1 + p_2)(p_3 + p_4) \right]\sigma^2+O(\sigma^3)\right\}
\label{L2}
\end{eqnarray}
where
\begin{align}
b_{1}&=(V_{01}^{12})^2,&b_{2}&= 26 (V_{11}^{11})^2-2 (X_{11}^{11})^2,&b_{3}&=(V_{01}^{12})^2 V_{11}^{11}, \nonumber\\
  b_{4}&= (V_{01}^{12})^4,&b_5&=18 (V_{02}^{12})^2.
\end{align}
To get the quartic term in the tachyon effective action on has to
subtruct the contribution from the tachyon in the propagator, that
corresponds to the $\sigma^0$ power -the constant term $1$- in
(\ref{L2}). Since, as already noticed, for a four point amplitude
only even powers of $\sigma$ need to be considered, one is left with
the coefficient of the $\sigma^2$ term in the sum. Performing the
integral over $\sigma$, one finally gets the formula for the quartic
term in the CSFT tachyon effective action (\ref{CSFT}) in the
time-dependent case
\begin{eqnarray}
A_4^{L=2}(p_1,p_2,p_3,p_4)=&&\lambda_c^2 g^2\int\prod_{i=1}^n(2\pi
dp_i)\delta(\sum_i p_i)
\phi(p_i)\frac{\lambda_c^{\frac{2}{3}(\sum_{i=1}^4 p_i^2 +p_1\cdot
p_2+p_3\cdot p_4)}}{1-(p_1+p_2)^2}\cr &&
\!\!\!\!\!\!\left[\frac{b_2}{4}+ b_3 p_1(p_2 - p_1) +b_4 p_2 p_4(p_2
- p_1)(p_4 - p_3)+b_5 p_2 p_4 \right] \label{L2quartic}
\end{eqnarray}

\end{document}